\documentclass[a4paper,11pt]{article}
\usepackage{pos}

\usepackage{amsfonts} 
\usepackage{amssymb} 
\usepackage{amsmath} 
\usepackage{graphicx} 
\usepackage{subfigure} 
\usepackage{array} 
\usepackage{dcolumn} 
\usepackage{bm} 
\usepackage{latexsym} 
\usepackage{longtable} 
\usepackage{hyperref} 
\usepackage{bm}
\usepackage{slashed}
\usepackage{bbold}
\usepackage{color}
\usepackage{multirow}
\usepackage{rotating}
\usepackage{comment}
\usepackage [english]{babel}
\usepackage [autostyle, english = american]{csquotes}
\MakeOuterQuote{"}
\usepackage{graphicx}
\usepackage{subfigure}
\usepackage{xcolor,cancel}
\usepackage{appendix}[toc,page]
\usepackage{setspace}
\usepackage{amsmath, amssymb}
\usepackage{slashed}

\newcommand{\be}{\begin{eqnarray}}
\newcommand{\ee}{\end{eqnarray}}

\newcommand{\kk}{\mathbf{k}}

\newcommand{\spiN}{\sigma_{\pi N}}

\title{Nucleon form factors and the pion-nucleon sigma term}
\ShortTitle{Form factors and Sigma term}

\author*[a]{Rajan Gupta}
\author[a]{Tanmoy Bhattacharya}
\author[a]{Vincenzo Cirigliano}
\author[b]{Martin Hoferichter}
\author[c]{Yong-Chull Jang}
\author[d]{Balint Joo}
\author[a]{Emanuele Mereghetti}
\author[a,e]{Santanu Mondal}
\author[a,g]{Sungwoo Park}
\author[g]{Frank Winter}
\author[h]{Boram Yoon}

\affiliation[a]{Los Alamos National Laboratory, Theoretical Division T-2, Los Alamos, NM 87545, USA}

\affiliation[b]{Albert Einstein Center for Fundamental Physics, Institute for Theoretical Physics, University of Bern, Sidlerstrasse 5, 3012 Bern, Switzerland}
\affiliation[c]{Physics Department, Columbia University, New York, NY 10027, USA}
\affiliation[d]{Oak Ridge Leadership Computing Facility, Oak Ridge National Laboratory, Oak Ridge, TN 37831, USA}

\affiliation[e]{Department of Physics and Astronomy, Michigan State University, MI, 48824, USA}
\affiliation[f]{Department of Computational Mathematics, Science and Engineering, Michigan State University, MI, 48824, USA}

\affiliation[g]{Jefferson Lab, 12000 Jefferson Avenue, Newport News, Virginia 23606, USA}
\affiliation[h]{Los Alamos National Laboratory, Computer Computational and Statistical Sciences Division, CCS-7, Los Alamos, NM 87545, USA}

\emailAdd{rajan@lanl.gov}

\abstract{This talk summarizes the progress made since Lattice 2021 in
  understanding and controlling the contributions of towers of
  multihadron excited states with mass gaps starting lower than of radial
  excitations, and in increasing our confidence in the extraction of
  ground state nucleon matrix elements. The most clear evidence for
  multihadron excited state contributions (ESC) is in
  axial/pseudoscalar form factors that are required to satisfy the PCAC relation
  between them. The talk examines the broader question--which and how many of
  the theoretically allowed positive parity states $N(\bm p)\pi(-\bm
  p)$, $N(\bm 0)\pi(\bm 0)\pi(\bm 0)$, $N(\bm p)\pi(\bm 0)$, $N(\bm
  0)\pi(\bm p),\ \ldots$ make significant contributions to a given nucleon 
  matrix element? New data for the axial, electric and magnetic form
  factors are presented. They continue to show trends observed in
  Ref.~\cite{Park:2021ypf}. The N${}^2$LO $\chi$PT analysis of the ESC
  to the pion-nucleon sigma term, $\spiN$, has been extended to
  include the $\Delta$ as an explicit degree of
  freedom~\cite{Gupta:2022aba}. The conclusion reached in
  Ref.~\cite{Gupta:2021ahb} that $N \pi$ and $N \pi \pi$ states each
  contribute about 10~MeV to $\spiN$, and the consistency between the
  lattice result with $N \pi$ state included and the phenomenological
  estimate is not changed by this improvement. 
  \looseness-1}

\FullConference{%
 The 39th International Symposium on Lattice Field Theory (LATTICE2022)
  8--13 August, 2022
  Bonn, Germany
}

\begin{document}
\maketitle
\section{Introduction}

The neutron and proton are stable bound states of quarks and gluons
whose structure is governed by Quantum Chromodynamics. (In this talk, 
weak decays, isospin breaking and electromagnetic
corrections are ignored.) Simulations of lattice QCD are being used to predict
their properties with increasing control over all systematic
uncertainties. In addition to controlling the three standard
systematics: extrapolation to the continuum ($a \to 0$) and infinite
volume ($L \to \infty$) limits and evaluating the results at
$M_\pi=135$~MeV, there are two additional interrelated challenges to
precision calculations of nucleon properties.  The first is the
exponential fall-off of the signal to noise ratio proportional to
$e^{-(M_N - 1.5M_\pi)\tau}$ in all nucleon correlation functions. In
the state-of-the-art calculations with $O(10^6)$ measurements on about
5000 configurations, a good statistical signal extends up to $\approx
2$~fm in 2-point correlation functions and up to $\approx 1.5$~fm in
3-point functions (see Ref.~\cite{Park:2021ypf} for background,
notation, methodology and description of lattices used).  The second
challenge is that at these source-sink separations, ESC are
significant even in the simplest observables such as nucleon charges
(see Fig.~\ref{fig:sigma}), and in some channels (such as the matrix
element of the fourth component of the axial current) they dominate
the signal~\cite{Jang:2019vkm,Park:2021ypf}.

\begin{figure}[h]  
    \includegraphics[width=0.32\linewidth]{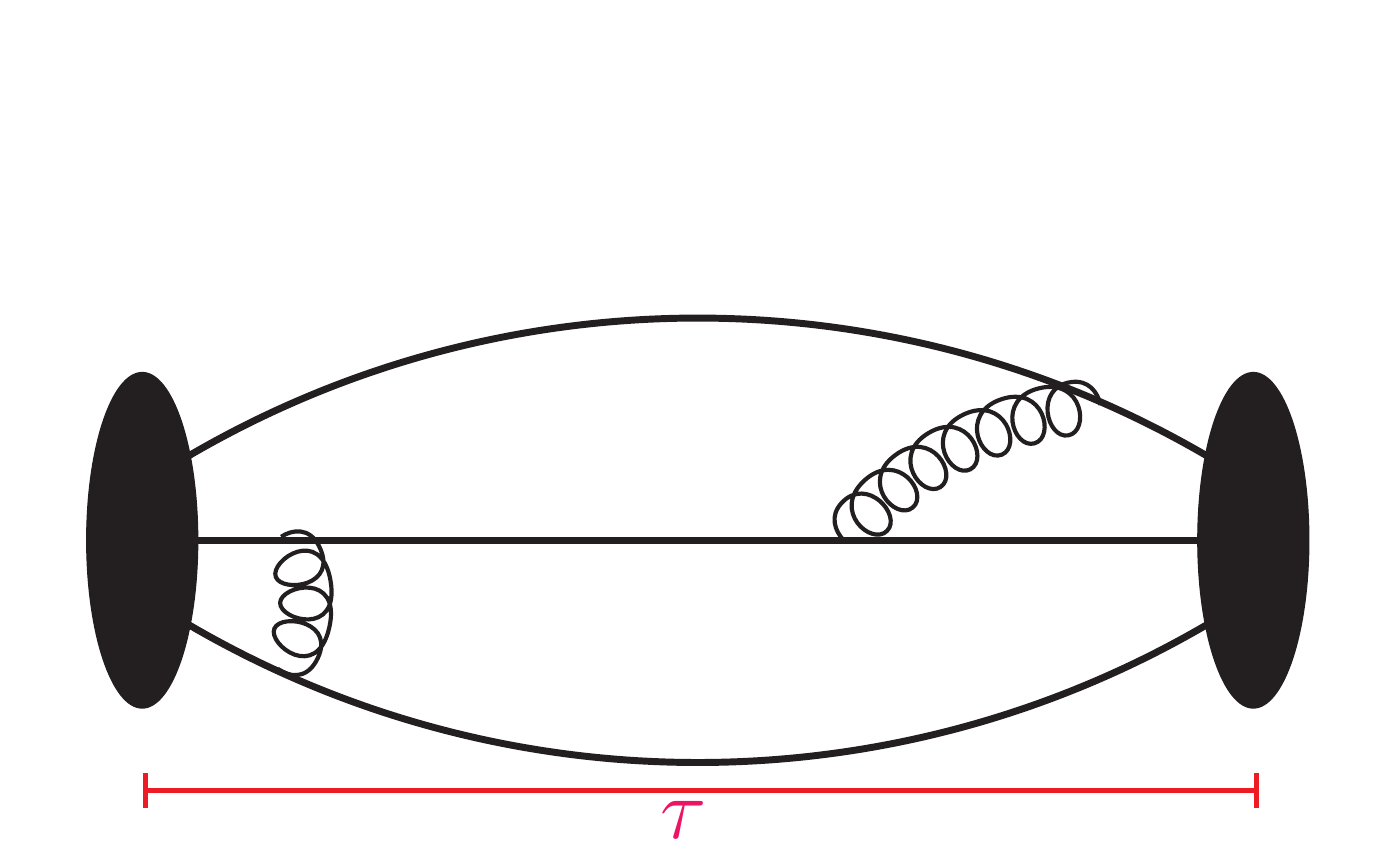}
    \includegraphics[width=0.32\linewidth]{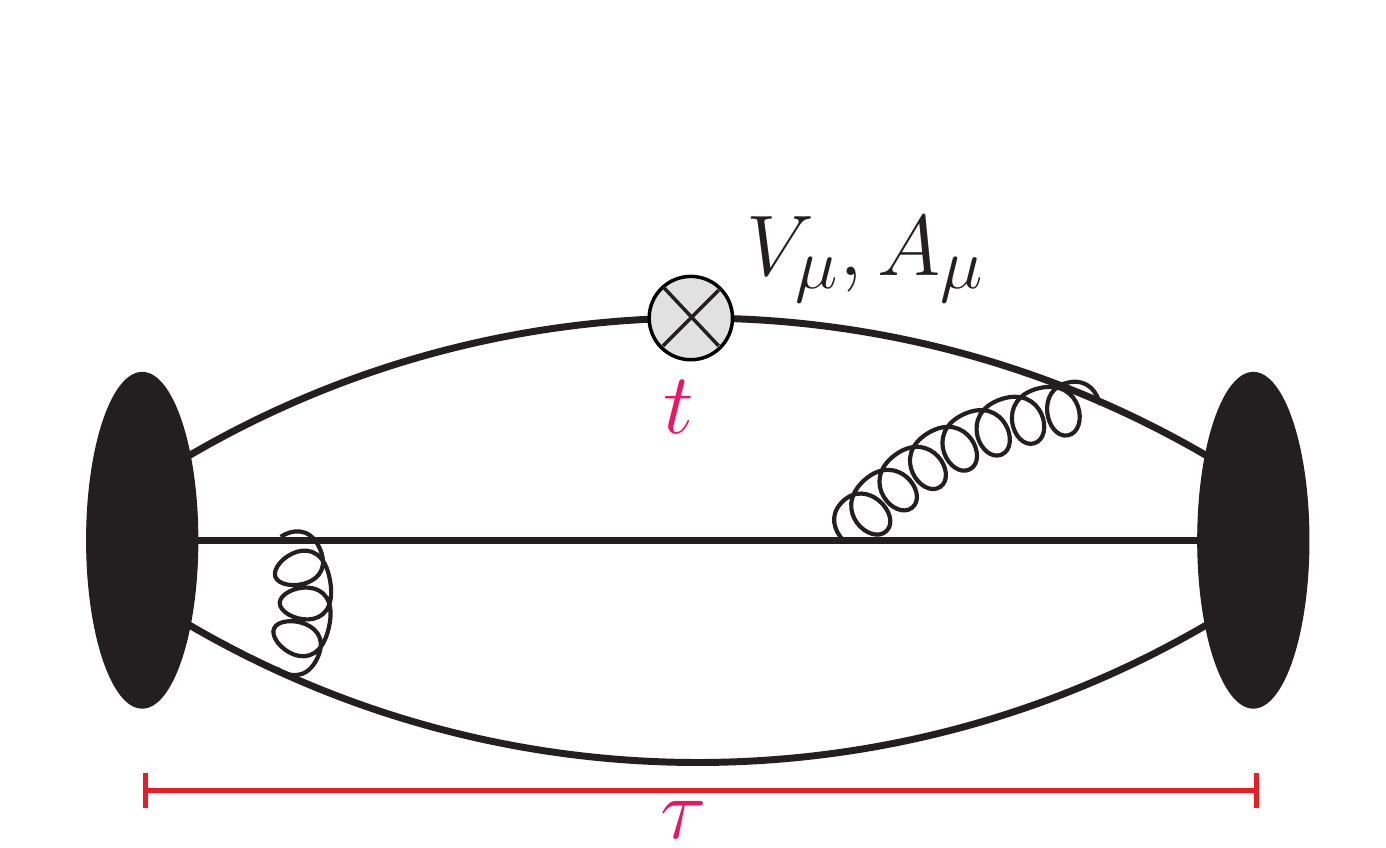}
    \includegraphics[width=0.32\linewidth]{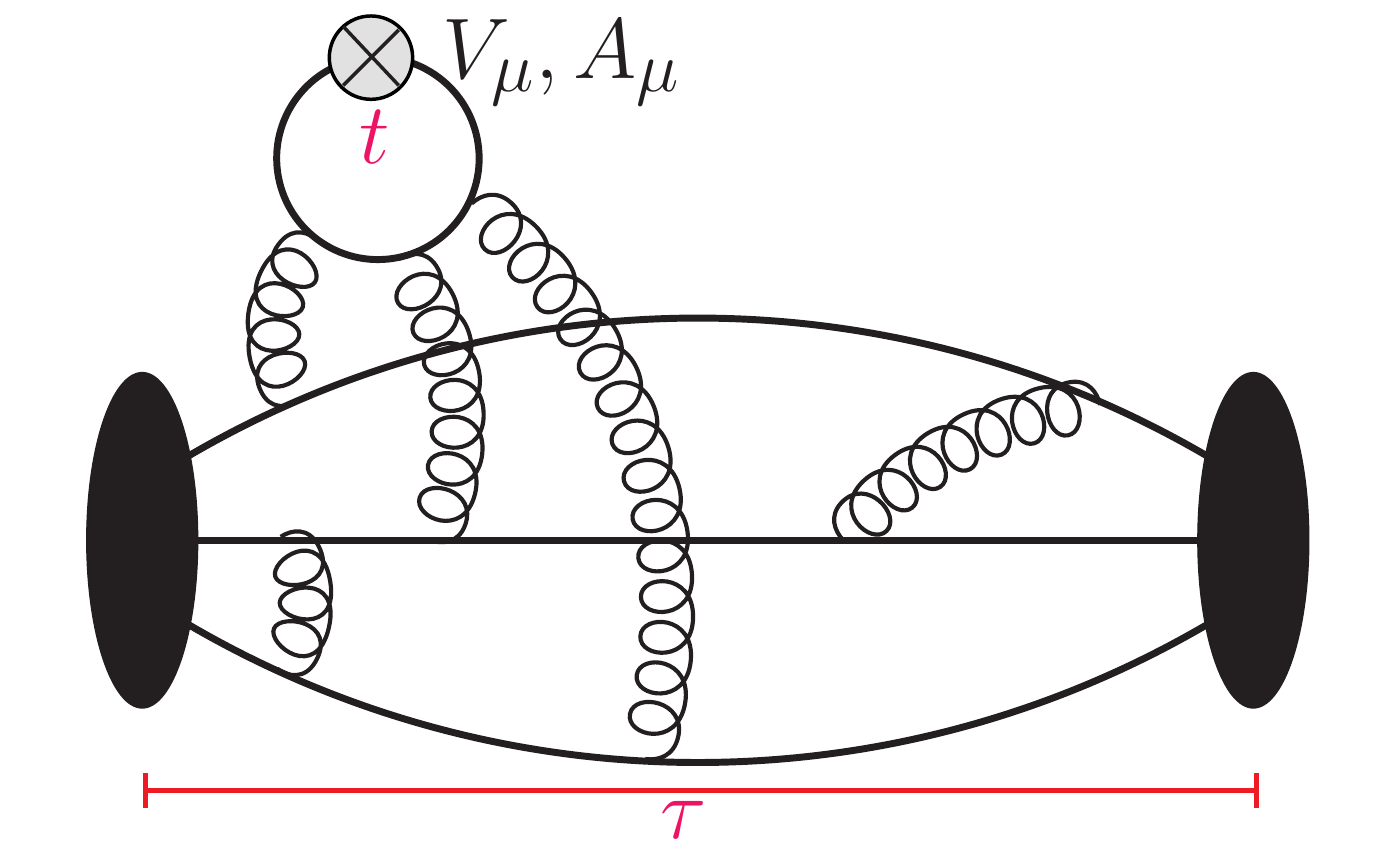}
\vspace{-0.1in}
\caption{Illustration of quark-line diagrams for 2-point (left),
  connected 3-point with the insertion of vector and axial operators
  $\bar u \gamma_\mu d$ and $\bar u \gamma_5 \gamma_\mu d$ (middle),
  and the additional disconnected contributions to matrix elements of
  flavor diagonal axial and vector operators $\bar q \gamma_\mu q$ and
  $\bar q \gamma_5 \gamma_\mu q$ (right). The operator, axial/vector
  current, is inserted at intermediate Euclidean time $t$ and with
  momentum $\vec q$. Each quark line represents the Feynman
  propagator, $S_F = {\cal D}^{-1}$ , given by the inverse of the
  Dirac matrix on that configuration. From these correlation
  functions, we extract the axial/vector form factors of the nucleon
  as outlined in Sec.~\ref{sec:FF}.\looseness-1}
\label{fig:2and3ptN}
\end{figure}

Theoretically, we know how to extract the matrix elements (ME) of various
operators within the ground state nucleon.  These are obtained by
making fits to the spectral decompositions of the correlation
functions that are the ensemble averages of quark-line diagrams shown
in Fig.~\ref{fig:2and3ptN}. The spectral decomposition of the 2- and
3-point correlation functions with the insertion of the current 
${\hat J}_\mu (= {\hat A}_\mu, \ {\hat V}_\mu,\ {\hat P},\ {\hat S},\ {\hat T}_{\mu\nu})$ at time $t$
  and with source-sink separation $\tau$ are given by
\begin{equation}
\Gamma^2_N = \sum_i |\langle \Omega |{\hat N} | N_i \rangle |^2 \ e^{-E_i \tau} ; \quad
\Gamma^3_N = \sum_{i,j} \langle \Omega |{\hat N} | N_i \rangle^\ast \ e^{-E_i (\tau-t) }
\langle N_i |{\hat J_\mu} | N_j \rangle     e^{-E_i t}   \langle N_j |{\hat N} | \Omega \rangle \,, 
\label{eq:SD2and3pion}
\end{equation}
from which we extract $\langle N_0 |{\hat A_\mu} | N_0 \rangle$ and
$\langle N_0 |{\hat V_\mu} | N_0 \rangle$ to calculate form factors.
The most direct strategy to get these ME is to fit $\Gamma^3_N$ and
resolve all the parameters. Even for a 2-state truncation of
Eq.~\eqref{eq:SD2and3pion}, this requires, in addition to $\langle N_0
|{\hat A_\mu} | N_0 \rangle$, resolving two energies, $E_{0,1}$, the
amplitude $A_0 \equiv \langle \Omega |{\hat N_i} | N \rangle$, and the not
subsequently used combinations: $A_{0}^\ast A_{1} \langle N_0 |{\hat
  A_\mu} | N_1 \rangle $, $A_{1}^\ast A_{0} \langle N_1 |{\hat A_\mu}
| N_0 \rangle $ and $A_{1}^\ast A_{1} \langle N_1 |{\hat A_\mu} | N_1
\rangle $.  Unfortunately, totally unconstrained fits to current data
at multiple values of $\{t,\tau\}$ are not sufficient to yield a
unique solution (large regions of parameter space give roughly the
same $\chi^2/dof$), so the resulting uncertainty in $\langle N_0
|{\hat A_\mu} | N_0 \rangle $ can be large.\looseness-1

In principle, all states of the transfer matrix with quantum numbers
of ${\hat N} $ contribute to the sums in
Eq.~\eqref{eq:SD2and3pion}. This set of states are the same for
$\Gamma^2_N$ and $\Gamma^3_N$. Thus, if we could take all the $E_i$
and $A_0$ from $\Gamma^2$, fits to $\Gamma^3$ would be greatly
improved and yield much better estimates for $\langle N_0 |{\hat
  A_\mu} | N_0 \rangle $. The challenge/question is---is the ordering
of the states by the size of their contribution to $\Gamma^2$ the same
as to $\Gamma^3$? The answer in many cases is NO. For many $\Gamma^3$,
contribution of towers of multihadron states, for example, $N(\bm
q)\pi(-\bm q), \forall {\bm q}\neq 0$ on the ${\bm p_j} = 0$ side in
$\Gamma^3$and $N(\bm q)\pi(-\bm {(p +q)})$ or $N(-\bm{(p+q})\pi(\bm
q), \forall {\bm q}$ on the ${\bm p_i} \neq 0$ side with ${\bm q} =
{\bm p}_j - {\bm p}_i$ contribute in addition to single-particle 
excited states. Such towers of states, labeled by relative
momenta $\bm q$, are typically not resolved in $\Gamma^2$ if a single
nucleon interpolating operator such as $\hat{N}(x) = \epsilon^{abc}
\left[ {q_1^a}^T(x) C \gamma_5 \frac{(1 \pm \gamma_4)}{2} q_2^b(x)
  \right] q_1^c(x) $ is used.\looseness-1

There are at least 5 states with positive parity and energy below
$N(1440)$ for our ensembles: $N$, ${N}(0,0,1) \pi(0,0,-1)$,
${N}(0,1,1) \pi(0,-1,-1)$, ${N}(0) \pi(0) \pi(0)$, and ${N}(0)
\pi(1) \pi(-1)$. This number grows as ${\vec q} \to 0$. To enlarge the
number of states included in the fit (order of truncation in
Eq.~\eqref{eq:SD2and3pion}) and yet nail the parameter space, one 
needs information from $\Gamma^2$ (at least $E_0$ and $A_0$) and physics inspired priors
with narrow width for the $E_i$.  On the output side, operationally, an n-state fit
function incorporates the influence of all states that give
significant contributions, so result of the fit for energies $E_1$ to $E_{n-1}$
are some combinations of the energies of all the excited states of
the transfer matrix.\looseness-1

This by itself is not a problem since we know approximately the energy
of the radial excitations ($N^{1/2}(1440), N^{1/2}(1710), \ldots$) and
of the non-interacting multihadron states, provided n-state fits with different
selections of these states are distinguished by the
$\chi^2/dof$ (i.e., a data driven analysis). For example, if the
$\chi^2/dof$ of a 2-state fit with say $E_1 = E_{N(1440)}$ is
significantly better than that with the lowest multihadron state with
positive parity ($E_1 = E_{N}(0,0,1) + E_\pi(0,0,-1) \sim 1230$~MeV),
then picking the result with $E_{N(1440)}$ is justified. The
problem we face is that the $\chi^2/dof$ of the fits to current data
are similar whereas the values of $\langle N_0 |{\hat A_\mu} | N_0
\rangle $ obtained are significantly different.  So one needs
additional information to pick between fits with different excited
states included. Here I describe two calculations for which ESC is
significant and additional information is needed to decide between the
fits. In the calculation of axial vector form factors it is satisfying
the axial Ward identity, i.e., PCAC, and in the calculation of the
pion-nucleon sigma term it is a $\chi PT$ analysis.\looseness-1

\section{A tempered surprise in the spectrum from the nucleon 2-point function}
\label{sec:2pt}

The spectrum of the transfer matrix in a finite box with nucleon
quantum numbers and lattice momenta $\bm p ={\bf n}2 \pi/La$ can be
determined from fits to the spectral decomposition of the two-point
function $\Gamma^{2}$ given in Eq.~\eqref{eq:SD2and3pion}. But 
how well do fits to $\Gamma^2$ with an interpolating
operator $\hat N$, which we tune to have minimum overlap with
excited states, capture the states that are significant for
$\Gamma^3$? 

\begin{figure}[htbp]  
\centering
\includegraphics[angle=0,width=0.24\textwidth]{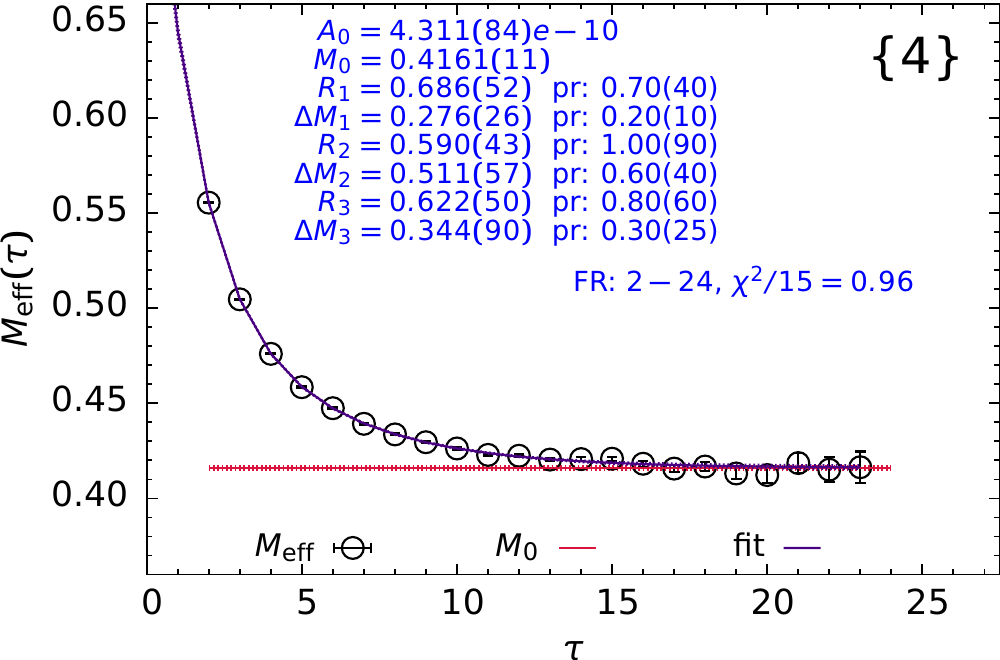}
\includegraphics[angle=0,width=0.24\textwidth]{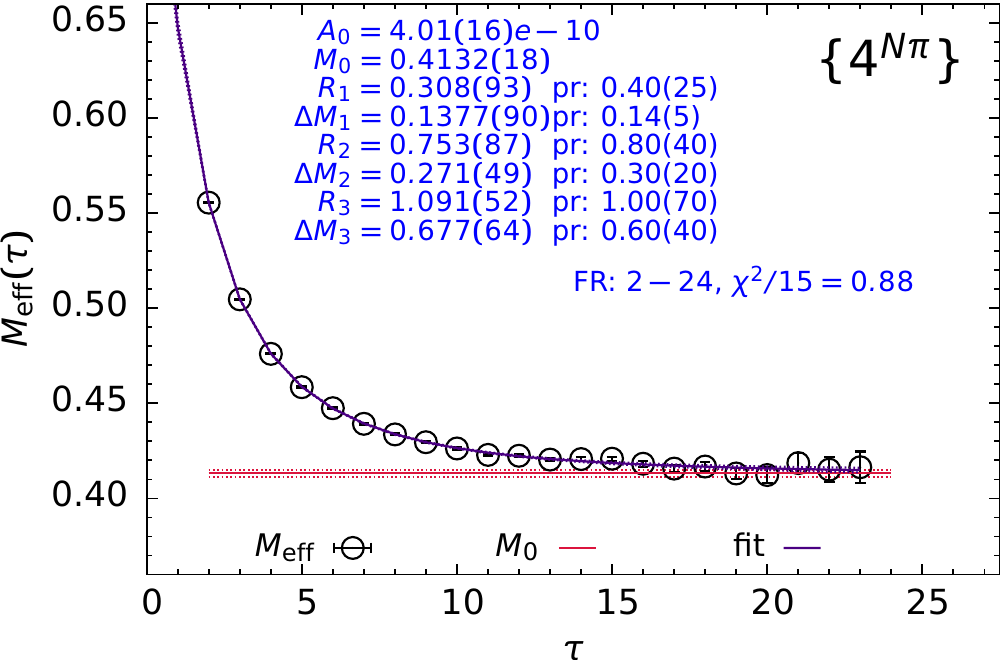}
\includegraphics[angle=0,width=0.24\textwidth,height=0.16\textwidth]{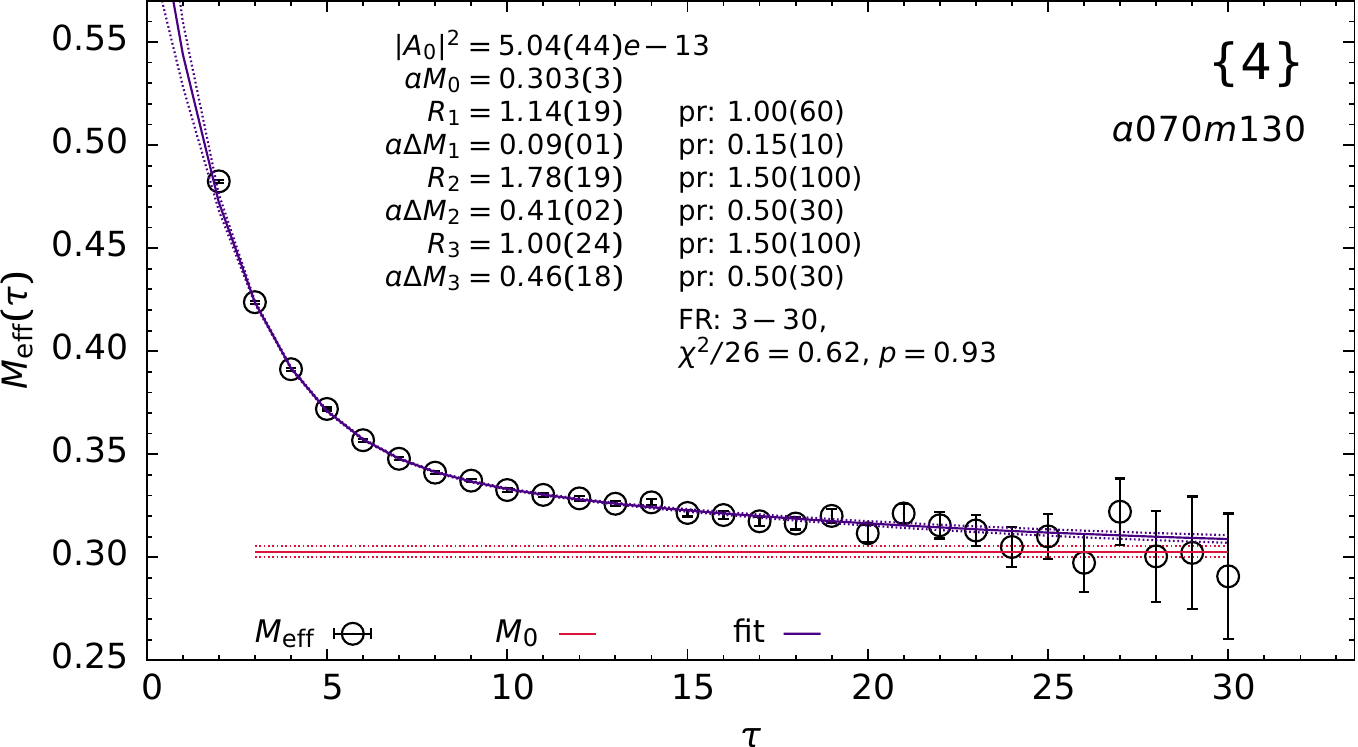}
\includegraphics[angle=0,width=0.24\textwidth,height=0.16\textwidth]{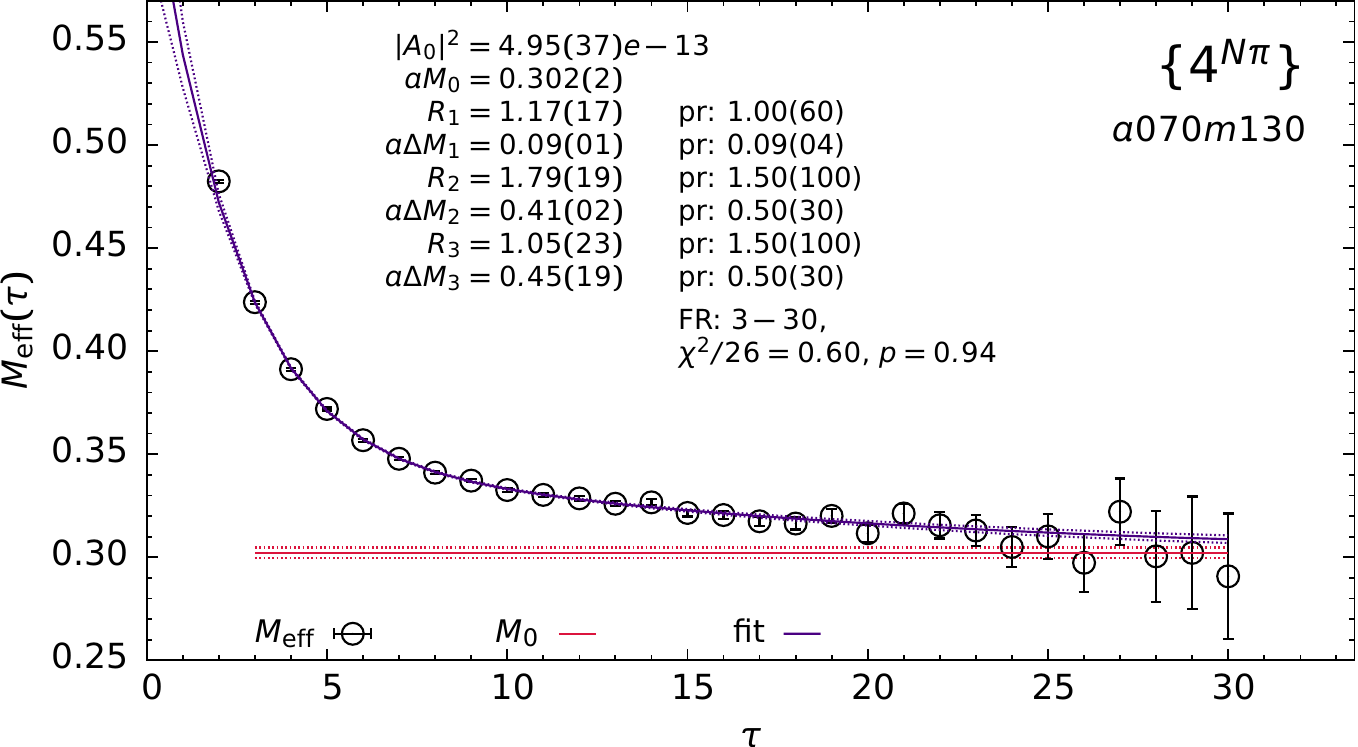}
\caption{Nucleon effective mass plots. The ``standard'' $\{4\}$ and
  the $\{4^{N\pi}\}$ fits to $a091m170$ ensemble data (left two
  panels) give different estimates of $E_i$ and $ A_i$. Fits to the
  $a070m130$ data give consistent estimates. }
\label{fig:2pt}
\end{figure}

An example of the conundrum of ESC with a single nucleon operator is
shown in Fig.~\ref{fig:2pt} (left 2 panels) of fits to
Eq.~\eqref{eq:SD2and3pion} truncated at four states using high
statistics $a091m170$ data. (See Ref.~\cite{Park:2021ypf} for
details.)  The left panel shows the standard analysis with wide priors
used only to stabilize the fit, while the second panel shows a fit
with a narrow prior for $E_1$ taken to be the energy of a
non-interacting $N (\bm 1) \pi(-\bm 1)$ state. The resulting $E_1$ are
about $ 1.5$ and 1.2~GeV, respectively. The two outcomes are not
distinguished by the augmented $\chi^2$ minimized in the fits. In
fact, in these 4-state fits there is a whole region of parameter space
that gives similar $\chi^2$ in which $E_1$ between $1.2 - 1.5$~GeV is equally
likely.  Furthermore, assuming $R_1 \equiv
|{\mathcal{A}}_1/{\mathcal{A}}_0|^2  \approx 1$, the contribution of a state
with $\Delta E_1 = 300$~MeV is still $20\%$ (5\%) at $\tau/a = 11$
(22), i.e., at source-sink separation $\tau$ of 1fm (2fm). \looseness-1

New data on the $a071m170$ and $a070m130$ ensembles break from the above 
pattern seen on the other 11 ensembles. Here, the ``standard'' $\{4\}$ and the $\{4^{N\pi}\}$ fits
give very similar spectrum ($E_i$ and $A_i$) with $E_1 \approx E_{N  \pi}$ 
and a large $E_2 > 2$~GeV (panels 3 and 4 inFig.~\ref{fig:2pt} for $a070m130$). 
The eventuality that different
initial points (priors) lead to the same minimum and expose the
$N \pi$ state in $\Gamma^2$ as $M_\pi \to 135$~MeV and $a \to 0$ is very
encouraging. Unfortunately, the $a070m130$ result is not stable, but
has depended on the statistics, leaving open the question--what
statistics will be needed to expose the multihadron states in $\Gamma^2$? Our current effort is to include 2 excited
states to analyze $\Gamma^3$ by trying different combinations of
possible states and input their $E_1$ and $E_2$ with physics driven
narrow priors. \looseness-1

\vspace{-0.6cm}

\section{Extraction of form factors}
\label{sec:FF}

\begin{figure*}[tbp] 
\subfigure
{
    \includegraphics[width=0.48\linewidth]{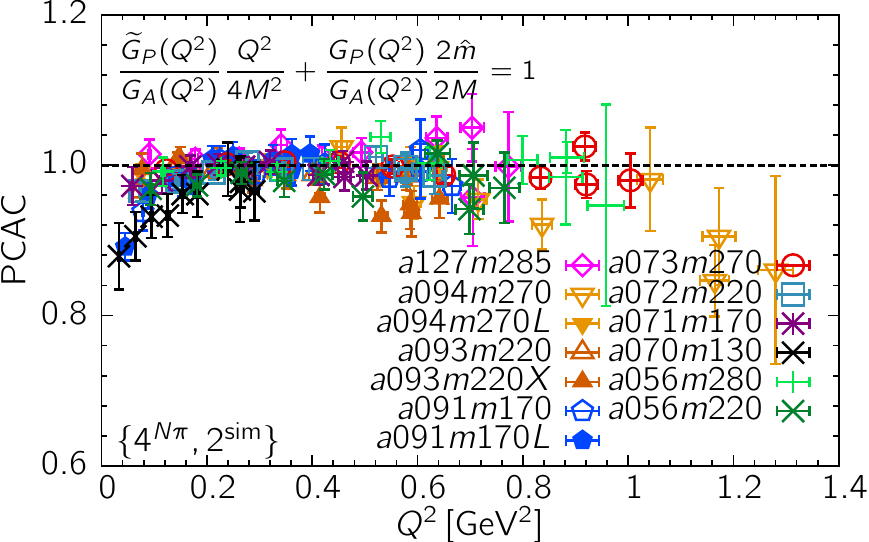}
\hspace{0.15cm}
    \includegraphics[width=0.50\linewidth]{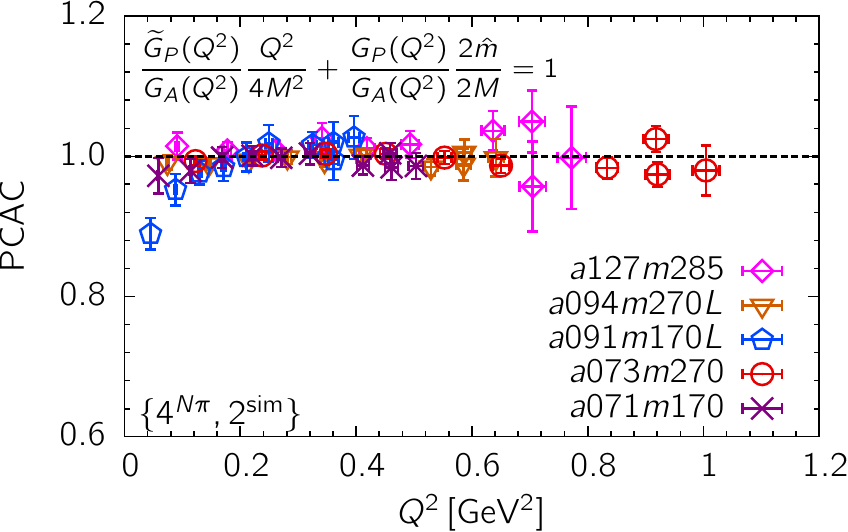}  
 }
 {
    \includegraphics[width=0.48\linewidth]{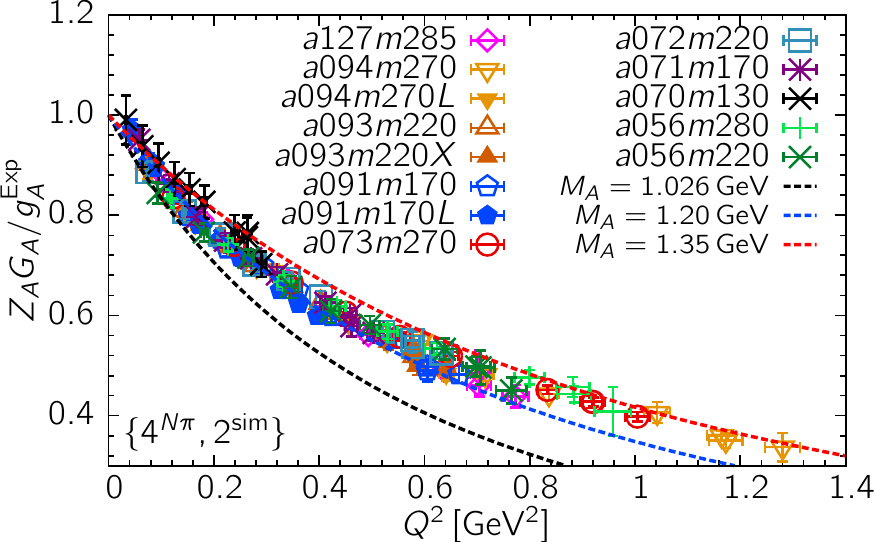}
    \includegraphics[width=0.50\linewidth]{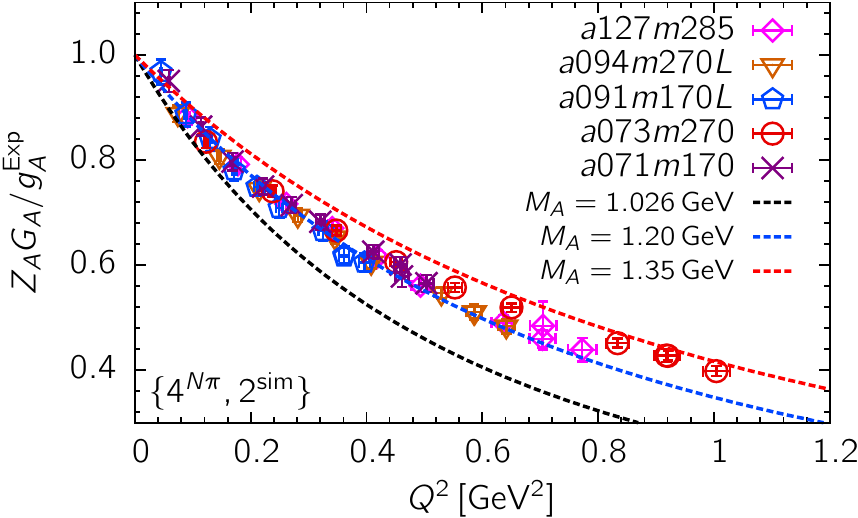} 
 }
 {   
    \includegraphics[width=0.48\linewidth]{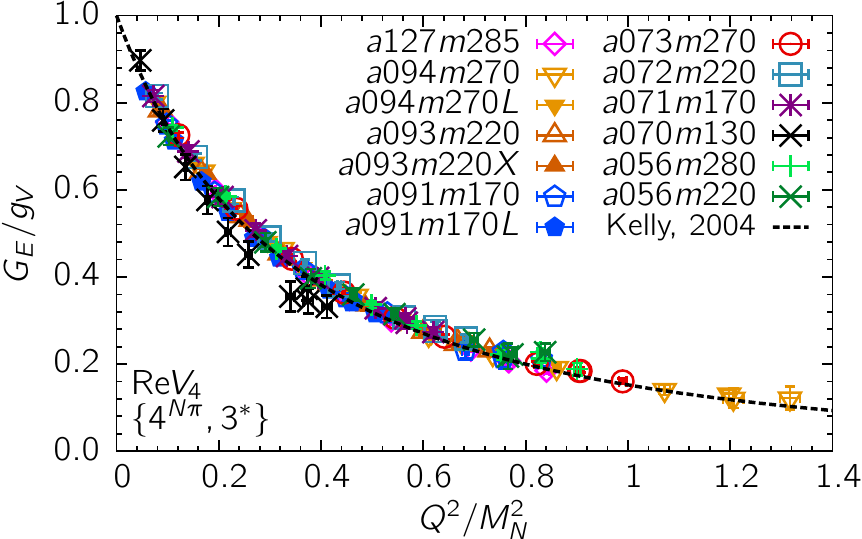} 
    \includegraphics[width=0.48\linewidth]{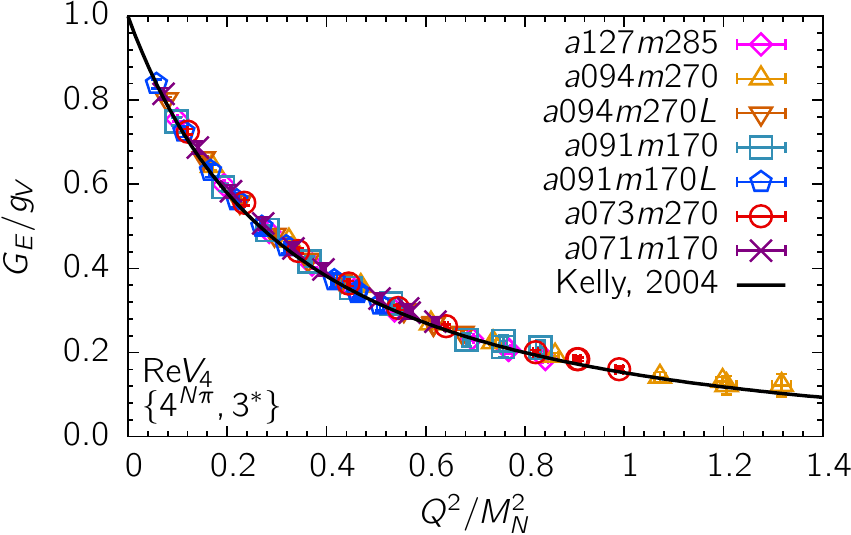} 
}
{
    \includegraphics[width=0.48\linewidth]{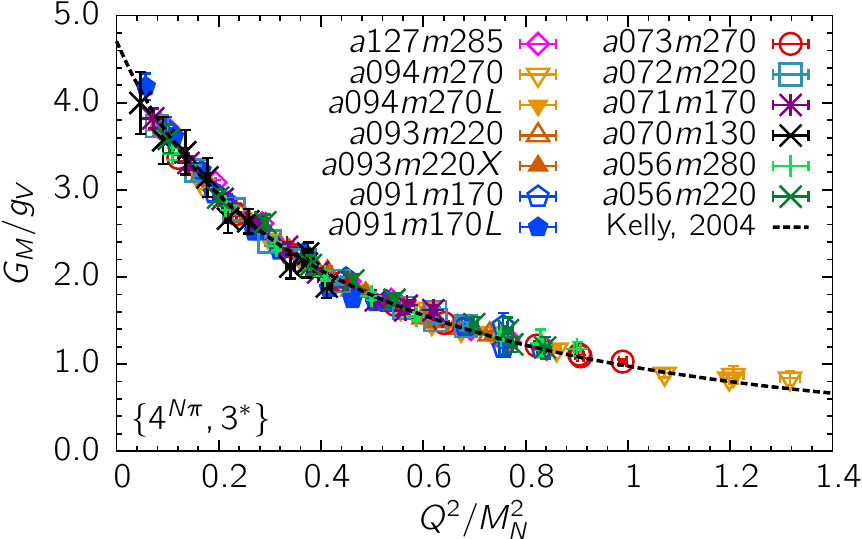}  
\hspace{0.3cm}
    \includegraphics[width=0.50\linewidth]{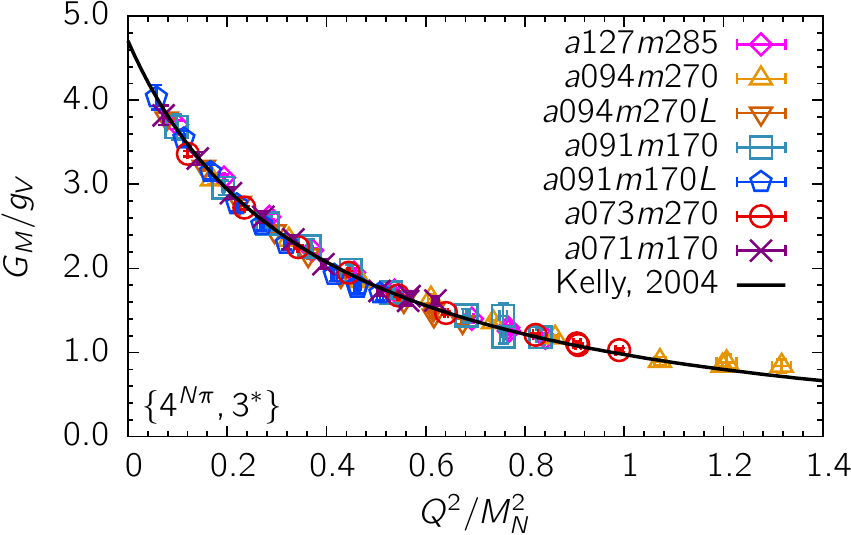}  
}
\vspace{-0.1cm}
\caption{Comparison of the preliminary data from 13 ensembles (left)
  with that published in Ref.~\protect\cite{Park:2021ypf} from 7
  ensembles (right).  The top row shows the degree to which the axial
  form factors satisfy the PCAC relation, Eq.~\protect\eqref{eq:PCAC}. 
  Rows 2-4 compare the axial $G_A(Q^2)$, the electric $G_E(Q^2)$,
  and the magnetic $G_M(Q^2)$ form factors. The statistics on the new 
  $M_\pi \approx 130$~MeV ensemble $a070m130$ are 40\% of the target 2500
  lattices. }
\label{fig:PCAC}
\end{figure*}

The Lorentz covariant decomposition of the iso-vector matrix elements (ME)
calculated within the nucleon ground state $| N({\bm p}_i,s_i)
\rangle$ extracted from the 3-point functions with the insertion of
the renormalized isovector axial, $A_\mu = Z_A {\overline u }\gamma_5
\gamma_\mu d$ and pseudoscalar $P = Z_P \overline{u} \gamma_5 d$
currents with momentum transfer $\vec q = \vec p_f - \vec p_i$ gives, 
in Euclidean space, the axial, $ G_A(Q^2)$, induced pseudoscalar, $
{\widetilde{G}_P(Q^2)}$, and the pseudoscalar, $ G_P(Q^2)$, form
factors:
\begin{align}
\label{eq:AFFdef}
\left\langle N({\bm p}_f,s_f) | A_\mu (\bm {q}) | N({\bm p}_i,s_i)\right\rangle  &= 
{\overline u}_N({\bm p}_f,s_f)\left( G_A(Q^2) \gamma_\mu
+ q_\mu \frac{\widetilde{G}_P(Q^2)}{2 M_N}\right) \gamma_5 u_N({\bm p}_i,s_i) \,, \nonumber \\
\left\langle N({\bm p}_f) | P (\bm{q}) | N({\bm p}_i)\right\rangle  &= 
{\overline u}_N({\bm p}_f) G_P(Q^2) \gamma_5 u_N({\bm p}_i) \,, 
\end{align}
where $Q^2 \equiv {\bf p}^2 - (E_f-E_i)^2 = -q^2$ is the spacelike
four-momentum squared. These three form factors must satisfy, up to
discretization errors, the following relation 
\begin{equation}
2 {\widehat m} G_P(Q^2) = 2 M_N G_A(Q^2) - \frac{Q^2}{2M_N} {\widetilde G}_P(Q^2) \,,
\label{eq:PCAC}
\end{equation}
that follows from the axial Ward identity, $\partial_\mu Z_A A_\mu - 2
Z_m { m} Z_P P=0$. Here ${\widehat m} \equiv Z_m Z_P (m_u +m_d)/(2
Z_A)$ is the average bare PCAC mass of the $u$ and $d$ quarks.  Result
of this test for the 2-state strategy $\{4^{N\pi},2^{\rm sim}\}$~\cite{Park:2021ypf}, shown in
the top row of Fig.~\ref{fig:PCAC},  indicates a growing deviation
as $Q^2 \to 0$, especially in the physical pion mass ensemble
$a070m130$. This suggests the need for a tuned $E_2$. 

Similarly, insertion of $V_\mu = Z_V {\overline u} \gamma_\mu d$, gives the 
Dirac $ F_1(q^2)$ and Pauli $ F_2(q^2)$ form factors:
\begin{equation}
\label{eq:VFF}
\left\langle N({\bm p}_f,s_f) | V_\mu^{\rm em} ({\bm q}) | N({\bm p}_i,s_i)\right\rangle  = 
{\overline u}_N({\bm p}_f,s_f)\left( F_1(Q^2) \gamma_\mu
+ \sigma_{\mu \nu} q_\nu
\frac{F_2(Q^2)}{2 M_N}\right)u_N({\bm p}_i,s_i),
\end{equation}
in terms of which the electric, $G_{E}$, and magnetic, $G_{M}$, 
form factors are given by  
\begin{align}\label{eq:sachs}
G_E(Q^2) &= F_1(Q^2) - \frac{Q^2}{4M_N^2}F_2(Q^2) \,, \\
G_M(Q^2) &= F_1(Q^2) + F_2(Q^2).
\end{align}
In addition to the PCAC constraint in Eq.~\eqref{eq:PCAC}, these form
factors have 4 experimental constraints: the conserved vector charge $g_V =
G_E|_{Q^2 = 0} = F_1|_{Q^2 = 0} =1$,
the difference of magnetic moments $(\mu^P - \mu^N) = G_M|_{Q^2 = 0}
= (F_1+F_2)|_{Q^2=0} = 4.7059$, the axial charge $g_A = G_A|_{Q^2 = 0}
= 1.276(2)$ from neutron $\beta$-decay, and ${\widetilde{G}_P|_{(Q^2 = 0.88m_\mu^2)}} = 8.06(55)$
from muon capture in the MuCap experiment~\cite{MuCap:2015boo}.  The 
goal is to determine the $Q^2$ behavior of these 5 form factors using 
lattice QCD with control over all sources of errors.

\vspace{-0.1cm}

\section{Progress in the calculation of form factors since Lattice 2021}
\label{sec:progress}

Our latest published results for the axial and vector form factors in
Ref.~\cite{Park:2021ypf} are based on seven 2+1-flavor Wilson-clover
ensembles, and these data were discussed at lattice 2021. Two of these
seven ensembles provided a finite volume study, thus the data were at
five distinct values of $\{a,M_\pi\}$.  This calculation has been
extended to 13 ensembles with 11 distinct values of $\{a,M_\pi\}$ with 
$M_\pi L > 4$. These data, especially the
physical pion mass ensemble $a070m130$ (only 40\% analyzed), should 
be considered preliminary. A summary of the highlights is
\begin{itemize}
\item
{\textbf{Axial form factor:}} In Ref.~\cite{Gupta:2017dwj}, we first
pointed out that the axial form factors obtained using the "standard"
method with excited state energies $E_i$ taken from $\Gamma^2$ fail to
satisfy the PCAC relation. The reason for this failure was identified
in Ref.~\cite{Jang:2019vkm}, and fits to the data for the 3-point
function $\langle \Omega | {\hat N}^\dagger A_4 {\hat N} | \Omega
\rangle$ confirmed the $\chi PT$ analysis~\cite{Bar:2018xyi} that the
$N \pi$ excited state contribution to the ME is enhanced and very
significant. Two analysis strategies were developed to include the
$N\pi$ state in accord with the pion-pole dominance hypothesis. In the
first labeled $\{4^{N \pi},3^\ast\}$, we input via a narrow prior the
non-interacting energy of the $N(0,0,1)\pi(0,0,-1)$ state for $E_1$ in
a 4-state fit to $\Gamma^2$ and use the output $E_i$ in a 3-state fit
to $\Gamma^3$. In the second labeled $\{4^{N \pi},2^{\rm sim}\}$, we
make a simultaneous fit to the five correlation functions with
insertions of $A_\mu$ and $P$ and take only $E_0$ and $A_0$ from
$\Gamma^2$. In $\{4^{N \pi},2^{\rm sim}\}$, the data for $A_4$ (which
are dominated by the excited state) drive the determination of $E_1$
that ends up being close to $N(0,0,1)\pi(0,0,-1)$, especially as
$M_\pi \to 135$~MeV. The advantage of the second method is---it is
totally data driven but the disadvantage is it is, so far, only
2-state (adding a third state brings in priors).  The comparison of
the new and old results with $\{4^{N \pi},2^{\rm sim}\}$ is presented
in the second row of Fig.~\ref{fig:PCAC}. The enlarged data set continues to
show mild dependence on $a$ and $M_\pi$.
\item
{\textbf{PCAC:}} Data in first row of Fig.~\ref{fig:PCAC}
show a significant ($\sim 12\%$) drop in the $a070m130$ data as $Q^2 \to 0$ 
compared to $\approx 50\%$ without $N \pi$ in the analysis~\cite{Gupta:2017dwj}. 
We are investigating whether this indicates the need for a second low
mass excited state added to the $\{4^{N \pi},2^{\rm sim}\}$
analysis. 
\item
{\textbf{Electric and magnetic form factors:}} the data (except for
the preliminary $a070m130$ data) in rows 3-4 of Fig.~\ref{fig:PCAC}
continue to show little dependence on $a$ and $M_\pi$ and agree with
the Kelly parameterization of the experimental data. Our best fit is
$\{4^{N \pi},3^\ast\}$ which includes a prior for $E_1$ that is close
to the $N \pi \pi $ state (on our lattices $E(N(0)\pi(0) \pi(0)) \approx
E(N(1)\pi(-1)$). The sensitivity of results to the value of $E_1$ is,
however, much smaller than in the axial case.  This is not
unreasonable since vector meson dominance suggests a coupling to the
$\rho$-meson (a $\pi \pi$ state), however, since it is much heavier
than the pion, the effect is likely to be smaller.  Chiral PT analysis
by B\"ar in Ref.~\cite{Bar:2021crj} indicates a $\approx 5\%$ effect
due to the pion loop. The $Q^2$ dependence and the magnitude of the
effect is consistent with the pattern seen in a summary of world
lattice data with the ``standard'' method shown in fig.~[22] in
Ref.~\cite{Jang:2019jkn}.  If the favorable situation shown here (EM form
factors have small systematics) persists, then increasing the
statistics and adding more ensembles is a viable strategy for
precision results in the near future.\looseness-1
\end{itemize}

\begin{figure}[h]  
\centering
    \includegraphics[width=0.92\linewidth]{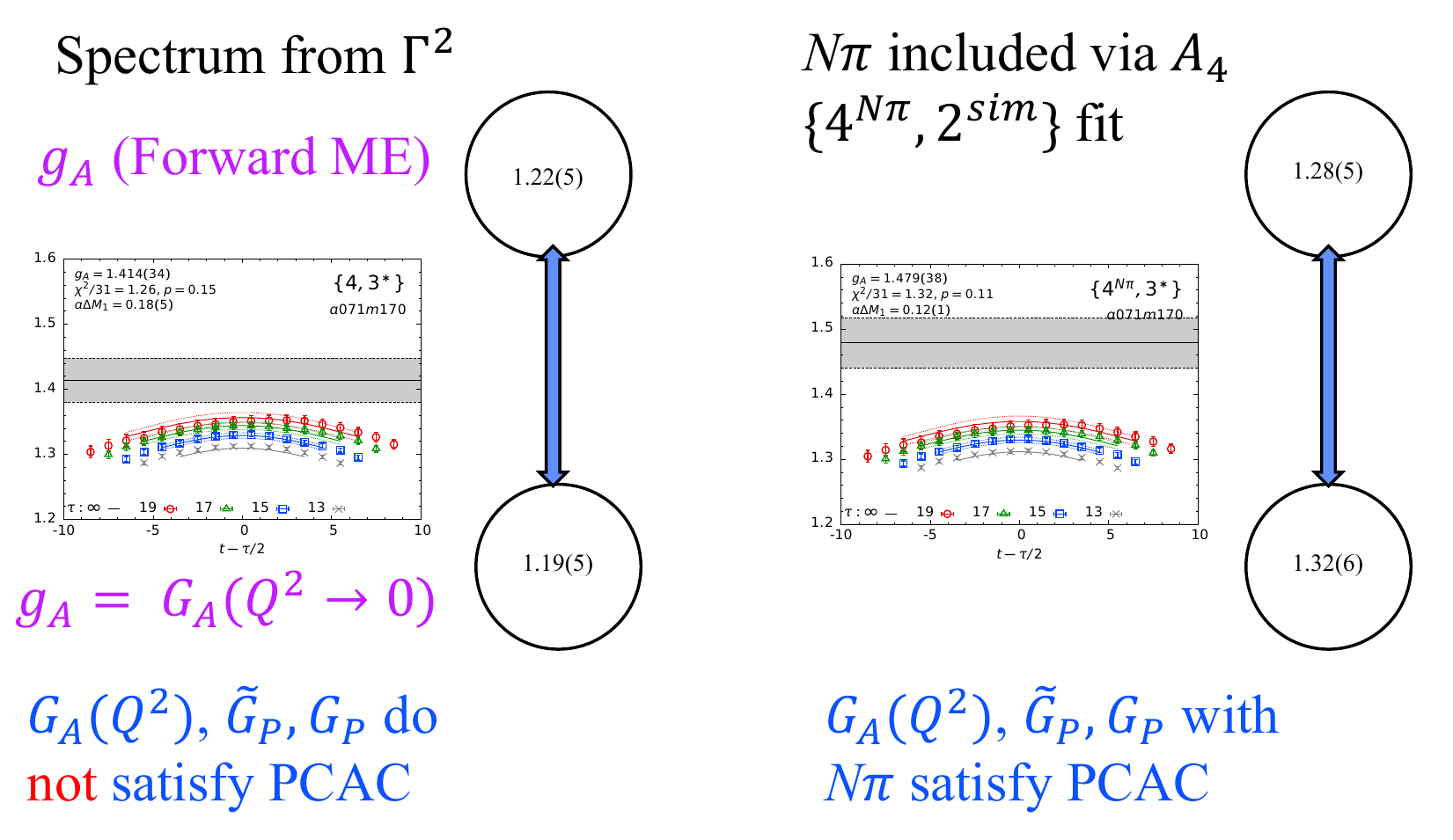}
\vspace{-0.1in}
\caption{Comparison of two strategies for extracting
$g_A^{u-d}$: (left) using the standard method, and (right) including 
 $N(0,0,1)\pi(0,0,-1)$ state in the analysis of both the forward matrix element
and of the form factors.   \looseness-1}
\label{fig:gAfig}
\end{figure}
\vspace{-0.2cm}

\section{Analysis of the isovector axial charge $g_A^{u-d}$}

The issue of which excited states need to be included in the analysis
of the charge $g_A^{u-d}$ is subtle (see~\cite{FlavourLatticeAveragingGroupFLAG:2021npn} 
for background). The enhanced effect of 
the $N(0,0,1)\pi(0,0,-1)$ state to the ME in the form factors
analysis, vanishes as ${\vec q} \to 0$, however, this state still
contributes at 1-loop in $\chi PT$ and could be a $\sim 5\% $
effect. In Fig.~\ref{fig:gAfig} we illustrate a consistency check
that indicates that the analysis including the $N(0,0,1)\pi(0,0,-1)$
state is required.  The figure illustrates a comparison of two
strategies for extracting $g_A^{u-d}$ with all the numbers taken from
Ref.~\cite{Park:2021ypf}.  On the left hand side is the ``standard''
analysis of $g_A^{u-d}$ extracted both from the forward matrix element
and from $G_A(Q^2 \to 0)$.  These two estimates must agree by
continuity! They do but the value is $\approx 5\%$ smaller than
experiment. Most important--the form factors {\it{\textbf{do not}}} satisfy PCAC.
The right part shows the $\{4^{N \pi},2^{\rm sim}\}$ analysis 
with $E_1 \approx E(N(0,0,1)\pi(0,0,-1))$. Again the two values
agree, are about 6\% larger than the ``standard'' value, still
consistent with experiment, and {\it{\textbf{the form factors satisfy PCAC}}}.  
Final analysis of all 13 ensembles, keeping in mind the 
discussion in Sec.~\ref{sec:2pt}, is underway!

\section{The pion-nucleon sigma term}

Flavor diagonal scalar charges of the nucleon, $g_S^{u,d,s,c}$, are 
important for many analyses such as 
the spin independent cross-section of dark matter with nuclear
targets~\cite{Bottino:2001dj,Hoferichter:2017olk}, lepton flavor violation in $\mu\to e$ conversion in
nuclei~\cite{Cirigliano:2009bz,Crivellin:2014cta}, and in electric
dipole
 moments~\cite{Engel:2013lsa,deVries:2016jox}. These are obtained from
 the forward matrix element of the scalar density $\bar{q} q$ (here it is implicit that the subtracted operator,
 $\bar{q} q - \langle \Omega |\bar{q} q |\rangle $, is used as discussed in Ref.~\cite{Bhattacharya:2005rb})
 between
 the nucleon ground state:\looseness-1
\begin{align}
g_S^q =  \langle N({\kk}=0,s)| \bar{q}  q | N({\kk}=0,s) \rangle.
\label{eq:gSdef}
\end{align}
The pion--nucleon $\sigma$-term $\spiN \equiv { m}_{ud}\, g_S^{u+d}
\equiv {m}_{ud} \, \langle N({\kk},s)| \bar{u} u + \bar{d} d |
N({\kk},s) \rangle$ is a fundamental parameter of QCD---it quantifies
the amount of the nucleon mass that comes from $u$- and $d$-quark
masses being non-zero.  In addition to lattice calculations, $\spiN$
has also been extracted phenomenologically from $\pi-N$ scattering via
the Cheng--Dashen low-energy theorem~\cite{Cheng:1970mx,Brown:1971pn}.

The FLAG 2021 report~\cite{FlavourLatticeAveragingGroupFLAG:2021npn}
highlights a tension between the lattice estimates that favor $\spiN
\approx 40$~MeV versus a value around $\spiN \approx 60$~MeV from
phenomenology~\cite{Hoferichter:2015dsa,RuizdeElvira:2017stg}.

\begin{figure}[tbp] 
\vspace{-0.1in}
    \centering
  \subfigure{
    \includegraphics[width=0.40\linewidth]{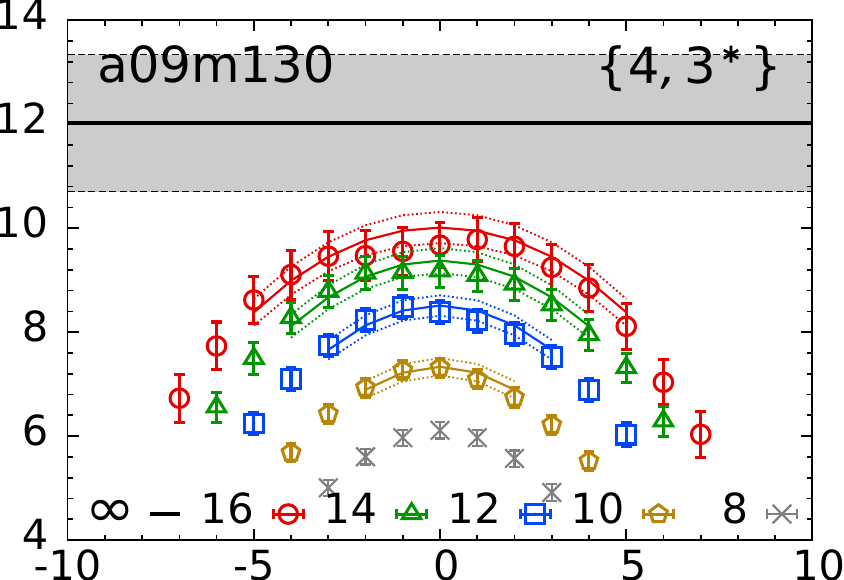} \hspace{0.5cm}
    \includegraphics[width=0.40\linewidth]{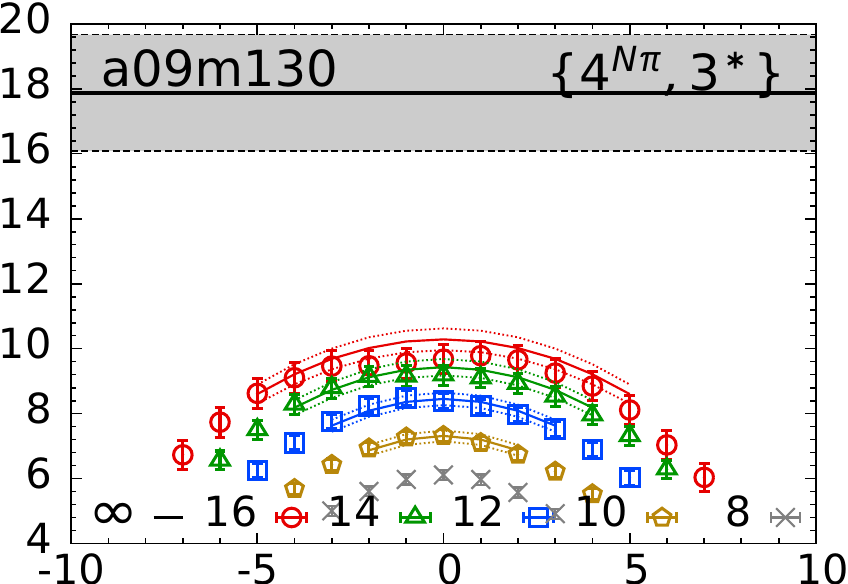} 
}
    \caption{The data for $g_S^{u+d+2l}$ from the physical mass,
      $a09m130$ ensemble and the two fits to remove ESC.  (Left)
      standard analysis and (Right) including $N \pi$ and $N \pi \pi $
      states.  Both panels are reproduced
      from~\protect\cite{Gupta:2021ahb}.  }
    \label{fig:sigma}
\end{figure}

The N${}^2$LO $\chi$PT analysis~\cite{Gupta:2021ahb} shows that there
is an enhanced contribution from $N \pi$ and $N \pi \pi $ states due
to the large coupling of the scalar source to two pions, i.e., a large
quark condensate~\cite{Gupta:2021ahb}, and a large disconnected
contribution. Including the $\Delta$ as an explicit state in the $\chi
PT$ analysis~\cite{Gupta:2022aba} does not change the conclusion
in~\cite{Gupta:2021ahb}. Our calculation, done in the isospin symmetric limit with
${m}_{ud} = (m_u + m_d)/2$, gives $\spiN \approx 40$~MeV with the
standard analysis (consistent with previous lattice estimates) while
the $\chi$PT motivated one, i.e., including contributions of $N \pi$ and $N
\pi \pi $ states, gives $\spiN \approx 60$~MeV, which is consistent
with phenomenology.
Our lattice analysis is very highly weighted by the data on the one
physical $M_\pi$ ensemble shown in
Fig.~\ref{fig:sigma}. This is expected as the difference in the $E_1$ used in
the two fits to remove ESC grows as $M_\pi \to 135$~MeV. Again, the two fits give different results but are not
differentiated by the $\chi^2$. To reach discrimination, our estimate
is a $\ge $10X increase in statistics to get similar precision at
$\tau = 18,\ 20$ as on the current $\tau = 14,\ 16$ data.  It is very important to confirm this exciting result 
on other physical pion mass ensembles.\looseness-1

\section{Conclusions}

While the lattice methodology for the extraction of matrix elements
from correlation functions using spectral decomposition is robust and
steady progress is being made in calculations, two related issues of the exponential
decay of the statistical signal with source-sink separation in all
nucleon n-point functions, and the contribution of low-lying
multihadron excited states have to be addressed before claiming robust, high
precision results. We have summarized our progress 
in the last year in the extraction of axial and electromagnetic form
factors and in the pion-nucleon sigma term.  Analyses with the inclusion of the
``$N\pi$'' state changes the results very significantly in both cases,
and we provide evidence that they should be included. Motivated by the deviation seen
in the PCAC relation in Fig.~\ref{fig:PCAC}, our current efforts are to
determine which second excited state,  gives the next largest
contribution and to include it in the fits to further improve estimates
of ground state matrix elements.\looseness-1

We conclude by mentioning related promising efforts to overcome
the signal to noise problem by analytic continuation of the
contour of integration (see
Refs.~\cite{Wagman:2016bam,Detmold:2021ulb}), and for ESC, the use of
a variational basis of nucleon interpolating operators that includes
multihadron operators to project on to the ground state of the nucleon
at much shorter source-sink separations~\cite{Barca:2022uhi}.

\acknowledgments
The calculations presented are based on two sets of ensembles: the
$2+1+1$-flavor HISQ ensembles generated by the MILC collaboration and
the $2+1$-flavor Wilson-clover ensembles generated by the
JLAB/W\&M/LANL/MIT collaborations. The calculations used the Chroma
software suite~\cite{Edwards:2004sx}. We gratefully acknowledge
computing resources provided by NERSC; OLCF at Oak Ridge through ALCC
awards NPH110 and LGT107, and INCITE awards PHY138 and HEP133; USQCD
clusters; and LANL Institutional Computing. Support for this work was
provided by the U.S. DOE Office of Science HEP and NP, and the SNSF. 
T. Bhattacharya and R. Gupta were partly
supported by the U.S. DOE, Office of Science, HEP under Contract
No. DE-AC52-06NA25396. T. Bhattacharya, R. Gupta, S. Mondal, S. Park,
and B. Yoon were partly supported by the LANL LDRD program, and
S. Park by the Center for Nonlinear Studies.

\bibliographystyle{JHEP}
\let\oldbibitem\bibitem
\def\bibitem#1\emph#2,{\oldbibitem#1}
\let\oldthebibliography\thebibliography
\renewcommand\thebibliography[1]{\oldthebibliography{#1}%
                                 \itemsep0pt\parskip0pt\relax}
\bibliography{ref}

\providecommand{\href}[2]{#2}\begingroup\raggedright\begin{thebibliography}{10}

\bibitem{Park:2021ypf}
S.~Park, R.~Gupta, B.~Yoon, S.~Mondal, T.~Bhattacharya, Y.-C. Jang et~al.,
  \emph{{Precision Nucleon Charges and Form Factors Using 2+1-flavor Lattice
  QCD}},  \href{https://arxiv.org/abs/2103.05599}{{\ttfamily 2103.05599}}.

\bibitem{Gupta:2022aba}
R.~Gupta, T.~Bhattacharya, M.~Hoferichter, E.~Mereghetti, S.~Park and B.~Yoon,
  \emph{{The pion-nucleon sigma term from Lattice QCD}},  in \emph{{10th
  International workshop on Chiral Dynamics}}, 3, 2022,
  \href{https://arxiv.org/abs/2203.13862}{{\ttfamily 2203.13862}}.

\bibitem{Gupta:2021ahb}
R.~Gupta, S.~Park, M.~Hoferichter, E.~Mereghetti, B.~Yoon and T.~Bhattacharya,
  \emph{{Pion\textendash{}Nucleon Sigma Term from Lattice QCD}},
  \href{https://doi.org/10.1103/PhysRevLett.127.242002}{\emph{Phys. Rev. Lett.}
  {\bfseries 127} (2021) 242002}
  [\href{https://arxiv.org/abs/2105.12095}{{\ttfamily 2105.12095}}].

\bibitem{Jang:2019vkm}
Y.-C. Jang, R.~Gupta, B.~Yoon and T.~Bhattacharya, \emph{{Axial Vector Form
  Factors from Lattice QCD that Satisfy the PCAC Relation}},
  \href{https://doi.org/10.1103/PhysRevLett.124.072002}{\emph{Phys. Rev. Lett.}
  {\bfseries 124} (2020) 072002}
  [\href{https://arxiv.org/abs/1905.06470}{{\ttfamily 1905.06470}}].

\bibitem{MuCap:2015boo}
{\scshape MuCap} collaboration, V.~A. Andreev et~al., \emph{{Measurement of the
  Formation Rate of Muonic Hydrogen Molecules}},
  \href{https://doi.org/10.1103/PhysRevC.91.055502}{\emph{Phys. Rev. C}
  {\bfseries 91} (2015) 055502}
  [\href{https://arxiv.org/abs/1502.00913}{{\ttfamily 1502.00913}}].

\bibitem{Gupta:2017dwj}
R.~Gupta, Y.-C. Jang, H.-W. Lin, B.~Yoon and T.~Bhattacharya, \emph{{Axial
  Vector Form Factors of the Nucleon from Lattice QCD}},
  \href{https://doi.org/10.1103/PhysRevD.96.114503}{\emph{Phys. Rev. D}
  {\bfseries 96} (2017) 114503}
  [\href{https://arxiv.org/abs/1705.06834}{{\ttfamily 1705.06834}}].

\bibitem{Bar:2018xyi}
O.~B{\"a}r, \emph{{$N\pi$-state contamination in lattice calculations of the
  nucleon axial form factors}},
  \href{https://doi.org/10.1103/PhysRevD.99.054506}{\emph{Phys. Rev. D}
  {\bfseries 99} (2019) 054506}
  [\href{https://arxiv.org/abs/1812.09191}{{\ttfamily 1812.09191}}].

\bibitem{Bar:2021crj}
O.~B{\" a}r and H.~{\v C}oli{\'c}, \emph{{N\ensuremath{\pi}}-state
  contamination in lattice calculations of the nucleon electromagnetic form
  factors}, \href{https://doi.org/10.1103/PhysRevD.103.114514}{\emph{Phys. Rev.
  D} {\bfseries 103} (2021) 114514}
  [\href{https://arxiv.org/abs/2104.00329}{{\ttfamily 2104.00329}}].

\bibitem{Jang:2019jkn}
Y.-C. Jang, R.~Gupta, H.-W. Lin, B.~Yoon and T.~Bhattacharya, \emph{{Nucleon
  electromagnetic form factors in the continuum limit from ( 2+1+1 )-flavor
  lattice QCD}}, \href{https://doi.org/10.1103/PhysRevD.101.014507}{\emph{Phys.
  Rev. D} {\bfseries 101} (2020) 014507}
  [\href{https://arxiv.org/abs/1906.07217}{{\ttfamily 1906.07217}}].

\bibitem{FlavourLatticeAveragingGroupFLAG:2021npn}
{\scshape Flavour Lattice Averaging Group (FLAG)} collaboration, Y.~Aoki
  et~al., \emph{{FLAG Review 2021}},
  \href{https://doi.org/10.1140/epjc/s10052-022-10536-1}{\emph{Eur. Phys. J. C}
  {\bfseries 82} (2022) 869}
  [\href{https://arxiv.org/abs/2111.09849}{{\ttfamily 2111.09849}}].

\bibitem{Bottino:2001dj}
A.~Bottino, F.~Donato, N.~Fornengo and S.~Scopel, \emph{{Size of the neutralino
  nucleon cross-section in the light of a new determination of the pion nucleon
  sigma term}},
  \href{https://doi.org/10.1016/S0927-6505(02)00107-X}{\emph{Astropart. Phys.}
  {\bfseries 18} (2002) 205}
  [\href{https://arxiv.org/abs/hep-ph/0111229}{{\ttfamily hep-ph/0111229}}].

\bibitem{Hoferichter:2017olk}
M.~Hoferichter, P.~Klos, J.~Men\'endez and A.~Schwenk, \emph{{Improved limits
  for Higgs-portal dark matter from LHC searches}},
  \href{https://doi.org/10.1103/PhysRevLett.119.181803}{\emph{Phys. Rev. Lett.}
  {\bfseries 119} (2017) 181803}
  [\href{https://arxiv.org/abs/1708.02245}{{\ttfamily 1708.02245}}].

\bibitem{Cirigliano:2009bz}
V.~Cirigliano, R.~Kitano, Y.~Okada and P.~Tuzon, \emph{{On the model
  discriminating power of mu ---\ensuremath{>} e conversion in nuclei}},
  \href{https://doi.org/10.1103/PhysRevD.80.013002}{\emph{Phys. Rev. D}
  {\bfseries 80} (2009) 013002}
  [\href{https://arxiv.org/abs/0904.0957}{{\ttfamily 0904.0957}}].

\bibitem{Crivellin:2014cta}
A.~Crivellin, M.~Hoferichter and M.~Procura, \emph{{Improved predictions for
  $\mu\to e$ conversion in nuclei and Higgs-induced lepton flavor violation}},
  \href{https://doi.org/10.1103/PhysRevD.89.093024}{\emph{Phys. Rev. D}
  {\bfseries 89} (2014) 093024}
  [\href{https://arxiv.org/abs/1404.7134}{{\ttfamily 1404.7134}}].

\bibitem{Engel:2013lsa}
J.~Engel, M.~J. Ramsey-Musolf and U.~van Kolck, \emph{{Electric Dipole Moments
  of Nucleons, Nuclei, and Atoms: The Standard Model and Beyond}},
  \href{https://doi.org/10.1016/j.ppnp.2013.03.003}{\emph{Prog. Part. Nucl.
  Phys.} {\bfseries 71} (2013) 21}.

\bibitem{deVries:2016jox}
J.~de~Vries, E.~Mereghetti, C.-Y. Seng and A.~Walker-Loud, \emph{{Lattice QCD
  spectroscopy for hadronic CP violation}},
  \href{https://doi.org/10.1016/j.physletb.2017.01.017}{\emph{Phys. Lett. B}
  {\bfseries 766} (2017) 254}
  [\href{https://arxiv.org/abs/1612.01567}{{\ttfamily 1612.01567}}].

\bibitem{Bhattacharya:2005rb}
T.~Bhattacharya, R.~Gupta, W.~Lee, S.~R. Sharpe and J.~M.~S. Wu,
  \emph{{Improved bilinears in lattice QCD with non-degenerate quarks}},
  \href{https://doi.org/10.1103/PhysRevD.73.034504}{\emph{Phys. Rev.}
  {\bfseries D73} (2006) 034504}
  [\href{https://arxiv.org/abs/hep-lat/0511014}{{\ttfamily hep-lat/0511014}}].

\bibitem{Cheng:1970mx}
T.~P. Cheng and R.~F. Dashen, \emph{{Is SU(2) x SU(2) a better symmetry than
  SU(3)?}}, \href{https://doi.org/10.1103/PhysRevLett.26.594}{\emph{Phys. Rev.
  Lett.} {\bfseries 26} (1971) 594}.

\bibitem{Brown:1971pn}
L.~S. Brown, W.~J. Pardee and R.~D. Peccei, \emph{{Adler-Weisberger theorem
  reexamined}}, \href{https://doi.org/10.1103/PhysRevD.4.2801}{\emph{Phys. Rev.
  D} {\bfseries 4} (1971) 2801}.

\bibitem{Hoferichter:2015dsa}
M.~Hoferichter, J.~Ruiz~de Elvira, B.~Kubis and {\relax Ulf-G}.~Mei\ss{}ner,
  \emph{{High-Precision Determination of the Pion-Nucleon {\(\sigma\)} Term
  from Roy-Steiner Equations}},
  \href{https://doi.org/10.1103/PhysRevLett.115.092301}{\emph{Phys. Rev. Lett.}
  {\bfseries 115} (2015) 092301}
  [\href{https://arxiv.org/abs/1506.04142}{{\ttfamily 1506.04142}}].

\bibitem{RuizdeElvira:2017stg}
J.~Ruiz~de Elvira, M.~Hoferichter, B.~Kubis and {\relax Ulf-G}.~Mei\ss{}ner,
  \emph{{Extracting the $\sigma$-term from low-energy pion-nucleon
  scattering}}, \href{https://doi.org/10.1088/1361-6471/aa9422}{\emph{J. Phys.
  G} {\bfseries 45} (2018) 024001}
  [\href{https://arxiv.org/abs/1706.01465}{{\ttfamily 1706.01465}}].

\bibitem{Wagman:2016bam}
M.~L. Wagman and M.~J. Savage, \emph{{Statistics of baryon correlation
  functions in lattice QCD}},
  \href{https://doi.org/10.1103/PhysRevD.96.114508}{\emph{Phys. Rev. D}
  {\bfseries 96} (2017) 114508}
  [\href{https://arxiv.org/abs/1611.07643}{{\ttfamily 1611.07643}}].

\bibitem{Detmold:2021ulb}
W.~Detmold, G.~Kanwar, H.~Lamm, M.~L. Wagman and N.~C. Warrington, \emph{{Path
  integral contour deformations for observables in $SU(N)$ gauge theory}},
  \href{https://doi.org/10.1103/PhysRevD.103.094517}{\emph{Phys. Rev. D}
  {\bfseries 103} (2021) 094517}
  [\href{https://arxiv.org/abs/2101.12668}{{\ttfamily 2101.12668}}].

\bibitem{Barca:2022uhi}
L.~Barca, G.~Bali and S.~Collins, \emph{{Towards $N$ to $N\pi$ matrix elements
  from Lattice QCD}},  \href{https://arxiv.org/abs/2211.12278}{{\ttfamily
  2211.12278}}.

\bibitem{Edwards:2004sx}
{\scshape SciDAC, LHPC, UKQCD} collaboration, R.~G. Edwards and B.~Jo{\'o},
  \emph{{The Chroma software system for lattice QCD}},
  \href{https://doi.org/10.1016/j.nuclphysbps.2004.11.254}{\emph{Nucl. Phys.
  Proc. Suppl.} {\bfseries 140} (2005) 832}
  [\href{https://arxiv.org/abs/hep-lat/0409003}{{\ttfamily hep-lat/0409003}}].

\end{thebibliography}\endgroup



\end{document}